\documentstyle[12pt,axodraw]{article}
\newcommand{\la}[1]{\label{#1}}
\newcommand{\be}{\begin{equation}}
\newcommand{\ee}{\end{equation}}
\newcommand{\ba}{\begin{eqnarray}}
\newcommand{\ea}{\end{eqnarray}}
\newcommand{\bi}{\begin{itemize}}
\newcommand{\ei}{\end{itemize}}
\newcommand{\rmi}[1]{{\mbox{\scriptsize #1}}}
\newcommand{\fig}{Fig.~}
\newcommand{\eq}{Eq.~}
\newcommand{\eqs}{Eqs.~}
\newcommand{\nr}[1]{(\ref{#1})}
\newcommand{\tr}{{\rm Tr\,}}
\newcommand{\nn}{\nonumber \\}
\newcommand{\fr}[2]{{\frac{#1}{#2}}}

\newcommand{\pdp}{\phi^\dagger\phi}
\renewcommand{\vec}[1]{{\bf #1}}

\def\lsi{\raise0.3ex\hbox{$<$\kern-0.75em\raise-1.1ex\hbox{$\sim$}}}
\def\gsi{\raise0.3ex\hbox{$>$\kern-0.75em\raise-1.1ex\hbox{$\sim$}}}
\newcommand{\lsim}{\mathop{\lsi}}
\newcommand{\gsim}{\mathop{\gsi}}

\begin{document}

\begin{titlepage}
\begin{flushright}
CERN-TH/98-281\\
NORDITA-98/56HE\\
hep-lat/9809004 
\end{flushright}
\begin{centering}
\vfill

{\bf THE ELECTROWEAK PHASE TRANSITION\\ IN A MAGNETIC FIELD}
\vspace{0.8cm}

K. Kajantie$^{\rm a,b,}$\footnote{keijo.kajantie@cern.ch},
M. Laine$^{\rm a,b,}$\footnote{mikko.laine@cern.ch},
J. Peisa$^{\rm c,}$\footnote{pyjanne@swansea.ac.uk},
K. Rummukainen$^{\rm d,}$\footnote{kari@nordita.dk} and
M. Shaposhnikov$^{\rm a,}$\footnote{mshaposh@nxth04.cern.ch} \\

\vspace{0.3cm}
{\em $^{\rm a}$Theory Division, CERN, CH-1211 Geneva 23,
Switzerland\\}
\vspace{0.3cm}
{\em $^{\rm b}$Department of Physics,
P.O.Box 9, 00014 University of Helsinki, Finland\\}
\vspace{0.3cm}
{\em $^{\rm c}$Department of Physics,
University of Wales Swansea, Singleton Park,\\
Swansea SA2 8PP, U.K.\\}
\vspace{0.3cm}
{\em $^{\rm d}$NORDITA, Blegdamsvej 17,
DK-2100 Copenhagen \O, Denmark}

\vspace{0.7cm}
{\bf Abstract}

\end{centering}

\vspace{0.3cm}\noindent
We study the finite temperature electroweak phase transition in an
external hypercharge U(1) magnetic field $H_Y$, using lattice Monte
Carlo simulations. For sufficiently small fields, $H_Y/T^2\lsim 0.3$,
the magnetic field makes the first order transition stronger, but it
still turns into a crossover for Higgs masses $m_H\lsim 80$ GeV. For
larger fields, we observe a mixed phase analogous to a type I superconductor, 
where a single macroscopic tube of the symmetric phase,
parallel to $H_Y$, penetrates through the broken phase. For the
magnetic fields and Higgs masses studied, we did not see 
indications of the expected Ambj{\o}rn-Olesen phase, which 
should be similar to a type II superconductor.

\vfill

\noindent
CERN-TH/98-281\\
NORDITA-98/56HE\\
November 1998

\vfill

\end{titlepage}

\section{Introduction}

Assume that there
exist non-vanishing magnetic fields in the Early Universe.
Their fate then depends strongly on their scale. 
Long-range magnetic fields (with a scale larger than
a few astronomical units today) are frozen in the primordial plasma
because of its high conductivity and may survive till the present time.
Magnetic fields on smaller scales dissipate by the time of
recombination and seem to leave no observational trace. Thus, direct
astronomical observations may put constraints on the existence
of magnetic fields on cosmological distances, but cannot provide any
bounds on the short-range fields. The boundary between the short-range
fields (which decay by the time $t$) and the
long-range fields (which still exist at time $t$) is given by the
critical size $l_0 \sim \sqrt{t/ \sigma} \sim ({1}/{T})
\sqrt{M_\rmi{Pl}/T}$, where $\sigma \sim T$ is the plasma
conductivity, $M_\rmi{Pl}$ is the Planck mass, 
and $T$ is the temperature of the Universe.

There were several attempts to explain the origin of the galactic
magnetic fields through the creation of so-called ``seed'' primordial fields
which may originally be weak but are amplified later through the
galactic dynamo mechanism. 
To get a non-negligible field-strength on the galaxy scale the
mechanism of magnetic field generation should be related to the
inflationary stage of the Universe expansion.
Generically, a whole spectrum of seed fields is produced, with an
amplitude which decreases with the length-scale. Thus, it may well
be that the Early Universe at temperatures higher than
the electroweak scale is filled with a stochastic (hyper)magnetic field,
whose contribution to the total energy density is 
not necessarily small at scales $\lsim l_0$
(for a review and references see, e.g., \cite{enqvist}).

Potentially, strong magnetic fields may influence different
processes in the Early Universe. Our primary interest here is in the
electroweak phase transition, the nature of which is essential for
electroweak baryogenesis. In spite of the expected stochastic and
space-dependent character of the magnetic field, a constant
homogeneous field approximation should be very good for this problem.
Indeed, 
the scale of surviving magnetic fields $l_0\sim 10^8/T$ at $T \sim 100$
GeV is much greater than the typical correlation lengths.

Without any external magnetic field, the electroweak phase
transition in the \linebreak SU(2)$\times$U(1) Minimal Standard Model 
is of the first order for small Higgs masses~\cite{su2u1}. 
The transition weakens with increasing $m_H$
so that the first order line has a second order endpoint~\cite{isthere}
of Ising type at $m_{H,c}\approx 72$ GeV~\cite{endpoint}. Beyond
that there is only a crossover. The two phases of the system,
the symmetric and the broken (or Higgs) phase, are thus analytically
connected.

If there is an external magnetic field, one expects
the transition to be significantly stronger~\cite{gs}. This
is simply because the hypercharge field $B$ contains a component of
the vector field $Z$, and $Z$ acts in a way similar to the magnetic
field in a superconductor: it vanishes in the broken Higgs phase.
Consequently, the broken phase has an extra contribution in the free
energy, and a system which would normally be considered to be deep in
the broken phase can now be in coexistence with the symmetric phase,
just like superconductivity can be destroyed by an external field.
Numerically, one obtains in the tree approximation that for a magnetic
field $H_Y/T_c^2\sim 0.5$, the electroweak phase transition would be of
the first order and strong enough for baryogenesis up to $m_H\sim 160$
GeV~\cite{gs}. A more precise 1-loop computation~\cite{eek}
(see also~\cite{others}) weakens the transition slightly:
for $H_Y/T_c^2\sim 0.3$, one seems to be able to go up to 
$m_H\sim 100$ GeV, while for larger fields, an instability may take place.

However, finite temperature perturbation theory cannot to be trusted in
the regime of large experimentally allowed Higgs masses, $m_H>80$ GeV.
Indeed, as mentioned, the first order electroweak
phase transition turns into a crossover for $m_H \lsim 80$
GeV~\cite{isthere,karschnpr,gurtler1,endpoint}
in the absence of a magnetic field, in contrast to the perturbative
prediction. Thus the effects of external magnetic fields should also be
studied non-perturbatively, and this is our objective here. We do
observe significant non-perturbative effects.

It is instructive to compare the present situation more 
precisely with  a superconductor (i.e., a U(1) gauge+Higgs theory) 
in an external magnetic field. This system 
has a very rich and well studied structure. There are two
possible responses to an imposed external flux:
\bi
\item type I, small $m_H$: the flux passes through a single domain;
\item type II, large $m_H$: the flux passes through a lattice of 
correlation length size vortices. In some cases, the vortex lattice 
can transform into a vortex liquid.
\ei
The fundamental difference between the two types 
is that for type II, the interface tension
between bulk symmetric and broken phases is negative so that it
is energetically favourable to split a single domain into a collection
of small subdomains, vortices.

The physics of (hypercharge) magnetic fields in SU(2)$\times$U(1)
theories has essential new features compared with superconductivity. 
First, the flux can now penetrate also the broken phase. Second, 
the broken phase massless gauge field $Q$ couples now to 
$W^\pm$ through a three-vertex, due to the non-Abelian structure 
of the theory. One may thus expect
qualitatively new phenomena. In fact, Ambj{\o}rn and Olesen \cite{ao}
have shown that for $m_H=m_Z$ the classical energy functional is minimized
by a configuration containing a $W^\pm$-condensate with a periodic
vortex-like structure. The question now is what happens in the full
quantum theory. 

The plan of the paper is the following. In Sec.~\ref{continuum} we
formulate the problem in continuum, and 
in Sec.~\ref{pert} we review briefly the perturbative
estimates. In Sec.~\ref{lattice} we describe how the system can be put
on the lattice. The numerical results are in Sec.~\ref{results}
and the conclusions in Sec.~\ref{conclusions}.

\section{Magnetic fields in the continuum}
\la{continuum}

The theory we consider is the effective 3d theory describing the finite
temperature electroweak phase transition in the Standard Model and in a
part of the parameter space of the MSSM. The Lagrangian is
\be
L_\rmi{3d}={1\over4} F_{ij}^aF_{ij}^a+{1\over4} B_{ij}B_{ij}+
(D_i\phi)^\dagger D_i\phi+m_3^2\phi^\dagger\phi+
\lambda_3(\phi^\dagger\phi)^2,
\label{contaction}
\ee
where $D_i=\partial_i+ig_3A_i+ig'_3B_i/2$ and
$B_{ij}=\partial_iB_j-\partial_jB_i$. The dynamics of the theory
depends on the three dimensionless parameters $x$, $y$ and $z$, defined
as
\be
x\equiv {\lambda_3\over g_3^2},\qquad
y\equiv {m_3^2(g_3^2)\over g_3^4},\qquad
z\equiv {g_3'^2\over g_3^2}.
\la{3dvariables}
\ee
These parameters can be expressed in terms of the underlying physical
4d parameters and the temperature; explicit derivations have been
carried out in~\cite{generic,mssm}. In the following, we fix $z=0.3$,
corresponding to $\sin^2\theta_W=0.23$.  The values 
$x\approx 0.10...0.13$ we have concentrated on, correspond
to Higgs masses $m_H\approx 72...82$ GeV in the Standard Model.
Moreover,
$H_Y^\rmi{3d}=H_Y^\rmi{4d}/\sqrt{T}$ up to $O(g'^2)$ corrections which
may be extracted from ref.~\cite{generic} and are not important
numerically. We neglect higher dimensional operators and assume that
the parameters of the theory are such that this is legitimate (the 
relevant requirements in the absence of a magnetic field are discussed
in \cite{generic}). 

The external magnetic field does contribute to
higher dimensional operators\footnote{We take here into account
also $H_Y$ when counting the dimension of an operator.} 
and thus the super-renormalizable 3d
Lagrangian in \eq\nr{contaction} is not valid 
for arbitrarily strong magnetic fields.
For instance, the 1-loop non-zero Matsubara
mode contribution to the scalar mass operator in an external field 
is of the form  $\lambda_3 g_3'^2 H_Y^2 \phi^\dagger\phi/(\pi T)^4$, 
and would give an $O(1)$ contribution to $y$ in \eq\nr{3dvariables}
for $H_Y \sim (\pi T)^2$.
Hence, one should add the condition $H_Y \ll (\pi T)^2$ to the list of
requirements for the validity of dimensional reduction.

We introduce the hypercharge magnetic field with an explicitly 
gauge-invariant \linebreak
method, which can thus be easily implemented on the
lattice\footnote{To fix a given electromagnetic field in the broken
phase, just tune the hypercharge field with the method described here
so that the desired value is obtained.}. 
We consider our system in a box with periodic boundary
conditions for all gauge-invariant operators. 
Then, the operator of the flux of the magnetic field through
a surface perpendicular to the $x_3$-axis,
\be
g_3' \Phi_B = \int dx_1dx_2 g_3' B_{12}(\vec{x}) =
g_3' \oint ds_i B_i(\vec{x}),
\la{contflux}
\ee
commutes with the Hamiltonian and is exactly conserved (the flux has been
multiplied by $g_3'$ to make it dimensionless also in 3d units). 
Thus, the equilibrium
thermodynamics of the system can be described by the density matrix
\be
\rho = Z^{-1} \exp(-H/T)\delta(\Phi_B-\Phi_B^\rmi{cl})
\la{microcan}
\ee
in the microcanonical formulation, where the value of the magnetic flux is
fixed by $\Phi_B^\rmi{cl}$. 

In the canonical formulation, the density matrix is given by
\be
\rho = Z^{-1} \exp[-(H - \mu_B\Phi_B)/T],
\ee
where $\mu_B$ is the ``chemical potential'' for the magnetic flux.
In more conventional notation, $\mu_B$ corresponds to an 
external field strength~$H_Y$, times the extent of the system in 
the $x_3$-direction, and $\Phi_B$ to a magnetisation density, 
times the area. 
The statistical sums in the two formulations are related, as usual, 
by the Legendre transformation.

It is seen from \eq\nr{contflux} that
in the microcanonical formulation which we use in practice, 
a magnetic field can be imposed with
suitable boundary conditions for $B_i(\vec{x})$. Although
$B_i(\vec{x})$ is not itself a physical (gauge-invariant) quantity, the
flux thus induced is. We stress that the local value of the
(hyper)magnetic field cannot be fixed as it is not an integral of
motion so that the dynamics of the system may prefer to distribute the
flux of the magnetic field in an inhomogeneous way.

To discuss the physical effects of a non-vanishing flux, we shall use
the following dimensionless combination: 
\be
b \equiv \frac{g_3'}{g_3^4} B_Y^\rmi{3d} \approx
\frac{g'}{g^4} \frac{B_Y^\rmi{4d}}{T^2} \approx 2.0
\frac{B_Y^\rmi{4d}}{T^2},
\la{b}
\ee
where $g_3' B_Y^\rmi{3d}\equiv g_3'\Phi_B/(\mbox{\rm area})$
is the magnetic flux density. For its Legendre transform, 
\be
H_Y^\rmi{3d} = \frac{\partial f(B_Y^\rmi{3d})}{\partial B_Y^\rmi{3d}},
\la{HYdef}
\ee
where $f(B_Y^\rmi{3d})$ is the free energy density obtained with the
density matrix in \eq\nr{microcan}, we use the corresponding notation
\be
h \equiv \frac{g_3'}{g_3^4} H_Y^\rmi{3d}. \la{h}
\ee
The effects of $b\neq 0, h\neq 0$ 
can be studied in perturbation theory. There are 
two types of effects: First, a magnetic field 
gives an extra contribution to the free
energy of the broken  phase relative to the symmetric phase, thus
changing the location of the critical curve. Second, a magnetic field 
affects the stability of homogeneous phases and can even lead to the 
emergence of completely new, inhomogeneous phases.

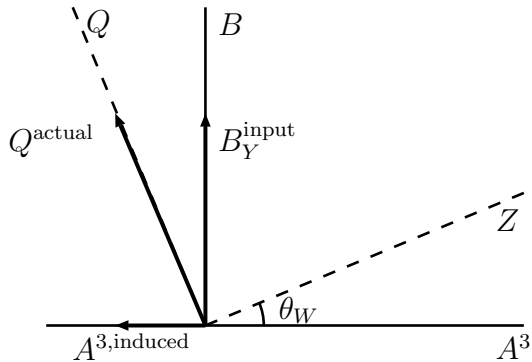
\begin{figure}[t]

\vspace*{-0.5cm}

\begin{center}
\begin{picture}(180,120)(0,0)

\SetWidth{1.0}
\Line(0,0)(180,0)
\Line(60,0)(60,120)
\DashLine(60,0)(180,50){5}
\DashLine(60,0)(10,120){5}
\CArc(60,0)(22,0,23)

\SetWidth{1.5}
\LongArrow(60,0)(26,0)
\LongArrow(60,0)(60,80)
\LongArrow(60,0)(26,80)

\Text(170,-8)[l]{$A^3$}
\Text(65,114)[l]{$B$}
\Text(170,40)[l]{$Z$}
\Text(15,114)[l]{$Q$}
\Text(88,6)[l]{$\theta_W$}
\Text(65,70)[l]{$B_Y^\rmi{input}$}
\Text(10,-8)[l]{$A^{3,\rmi{induced}}$}
\Text(2,70)[5]{$Q^\rmi{actual}$}
\end{picture}

\end{center}
\caption[a]{
An illustration of how the electromagnetic field $Q$ in the broken
phase is related to the external hypermagnetic field $B_Y$:
$Q^\rmi{actual}=B_Y^\rmi{input}\cos^{-1}\theta_W$.
$A^{3,\rmi{induced}}$ is such that the projection to the $Z$-direction
vanishes.} \la{fields} \end{figure}

\section{The phase structure in perturbation theory \la{pert}}

{\em Homogeneous phases.} 
Let us first consider the case of a fixed $b$, and assume that the 
ground states of the system are homogeneous. 
Then there is an extra free energy density related to the magnetic field.
In the symmetric phase, the non-Abelian field
$A^a_i$ does not have an expectation value, and the free-energy density
related to gauge fields is just $B_Y^{2}/2$. In the broken phase, it is
the electromagnetic field $Q$ which can have an expectation value, and
according to \fig\ref{fields}, its contribution is $B_Y^2/(2
\cos^2\theta_W)$. Hence, the phase equilibrium condition for the
effective potential $V(\phi,y)$ is
\be
V(0,y_c) +\fr12 (B_Y^\rmi{3d})^2= 
V(v,y_c) +\fr12 (B_Y^\rmi{3d})^2 \cos^{-2}\theta_W,
\la{treel}
\ee
where $v$ is the location of the broken minimum. It follows that for
$b>0$, there is at tree-level a first order transition 
between homogeneous phases along the
critical curve $y_c = - b \sqrt{2x}$, with the broken phase vacuum 
expectation value $v^2/g_3^2= b \sqrt{{2}/{x}}$.

{\em Mixed phases.}
However, as is usual, one has to consider 
also mixed phases in addition to homogeneous 
phases when working with the micro\-canoni\-cal ensemble\footnote{We thank
M. Tsypin for pointing out to us the importance of this consideration.}. 
A mixed phase consists of macroscopic domains of the symmetric and broken 
phases, and the magnetic flux can be distributed unevenly between them. 
Mixed phases do not appear in the canonical formulation, which thus
leads to a simpler phase diagram. Let us, therefore, see what happens
if the canonical variable $h$ is fixed instead of $b$.

Note first that according to \eq\nr{HYdef}, 
\be
H_Y^\rmi{3d} = B_Y^\rmi{3d} \quad (\mbox{symmetric phase}), \quad
H_Y^\rmi{3d} = B_Y^\rmi{3d} \cos^{-2}\theta_W \quad (\mbox{broken phase}).
\la{hbrel}
\ee
Making the Legendre transformations,
the condition in \eq\nr{treel} is replaced by
\be
V(0,y_c) -\fr12 (H_Y^\rmi{3d})^2= 
V(v,y_c) -\fr12 (H_Y^\rmi{3d})^2 \cos^{2}\theta_W.
\la{Htreel}
\ee
This leads to $y_c=-h\sqrt{2x/(1+z)}$. One can now go back to 
the microcanonical ensemble, to find out that for fixed $b$,
the canonical transition corresponds 
to a band with a finite width, 
$y=y_\rmi{min}...y_\rmi{max}$, where
\be
y_\rmi{min}= 
-b\sqrt{2x}\cos^{-1}\theta_W,\quad
y_\rmi{max}= 
-b\sqrt{2x}\cos\theta_W, \la{binter}
\ee 
and $\cos\theta_W=(1+z)^{-1/2}$.
Thus, the single transition line obtained from \eq\nr{treel} 
splits, in fact, into two transitions
between which the ground state is the mixed phase. 
The same result can, of course, also be obtained directly
in the microcanonical formulation, by minimizing the bulk 
free energy of a system with certain volume fractions of
the symmetric and broken phases, and an uneven distribution
of the flux.

\begin{figure}[t]


\begin{minipage}[c]{8cm}
\epsfysize=7cm\epsfbox{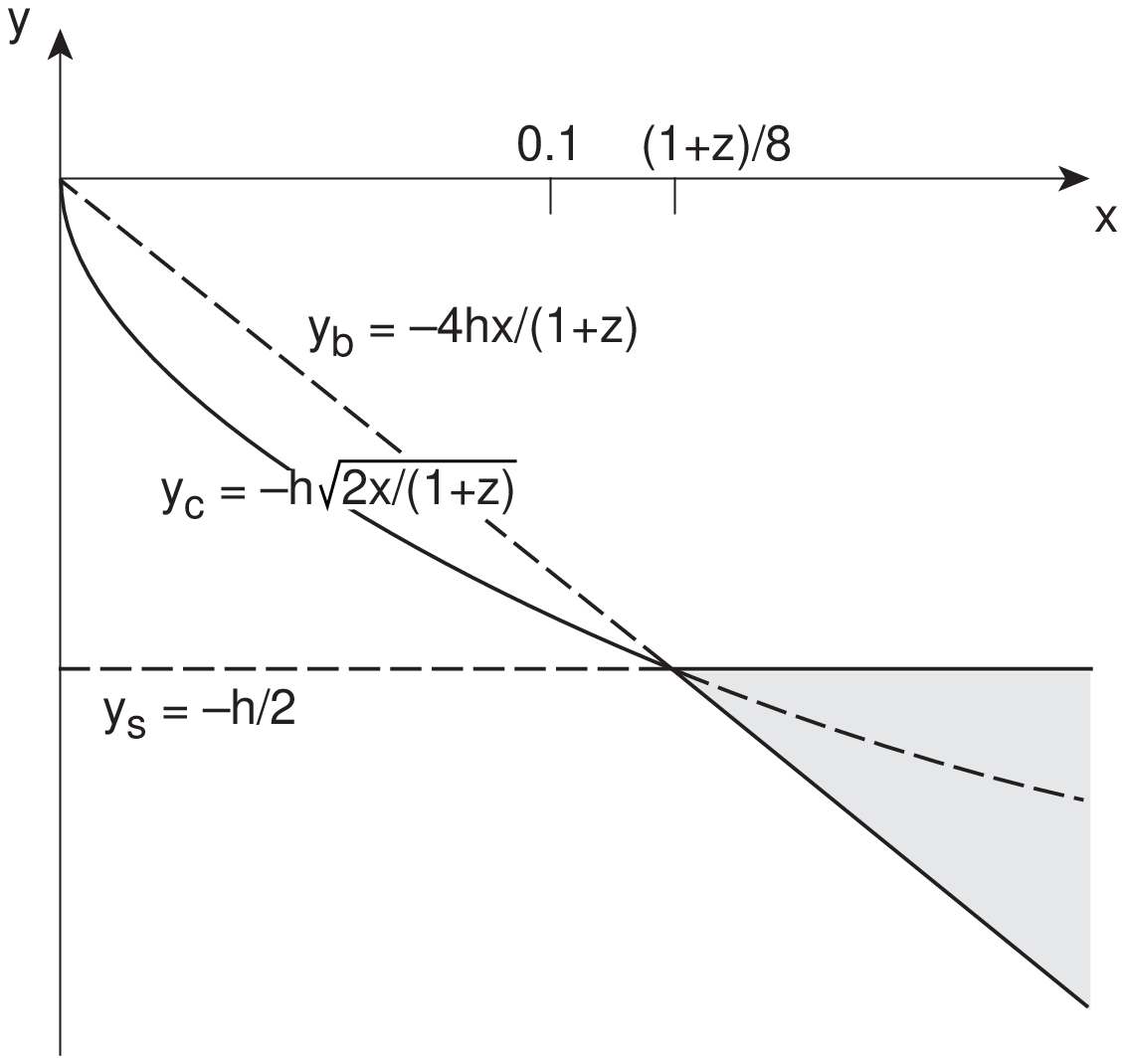}
\end{minipage}%
\hspace*{0.4cm}%
\begin{minipage}[c]{8cm}

\vspace*{1.2cm}

\epsfysize=7.5cm\epsfbox{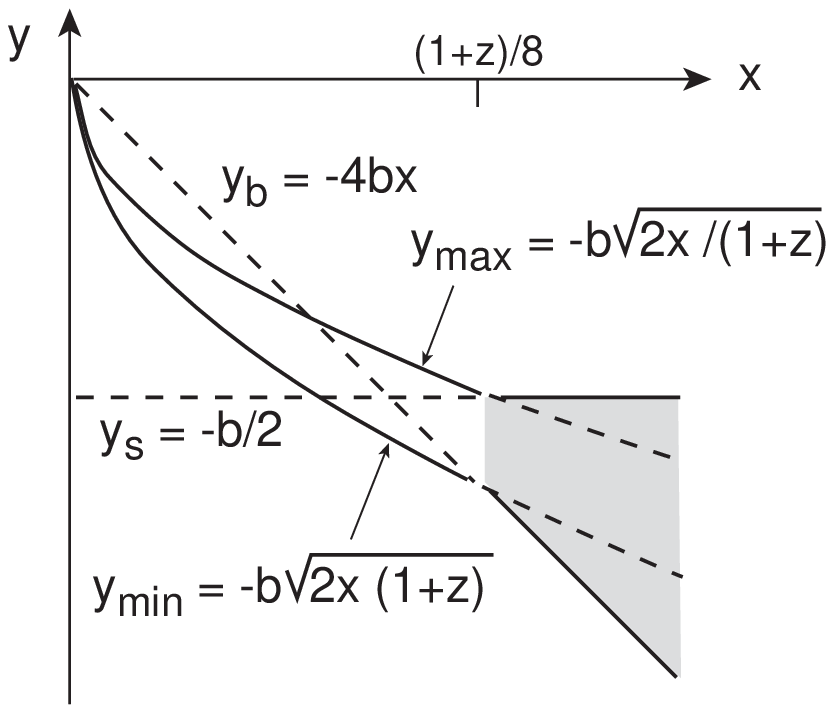} 
\end{minipage}

\caption[a]{The tree-level phase structure in an
external field. The dimensionless variables appearing
are defined in \eqs\nr{3dvariables}, \nr{b}, \nr{h}. 
{\em Left:} For a fixed field strength $h$, there is 
a first order transition along $y=y_c$ and
an instability within the shaded region.
{\em Right:} For a fixed flux density $b$, 
the first order transition corresponds 
to a band $y=y_\rmi{min}...y_\rmi{max}$ of a mixed 
phase.}
\la{treelevel}
\end{figure}

Finally, it should be noted that a mixed phase can 
only appear in a large total volume~$V$, when 
the free energy cost of the additional surfaces
is less than the free energy gain obtained by going 
from a homogeneous phase to a mixed phase.
For instance, at the point  $y=-b\sqrt{2x}$, derived from \eq\nr{treel} 
and lying close to the middle of the interval in \eq\nr{binter},
surfaces can only appear if
\be
\fr12 b^2 \tan^2(\theta_W/2) (V g_3^6)  > A \sigma, \la{finvol}
\ee
where $A$ is the area of the surfaces,
$\sigma$ is their surface tension, 
and $\tan^2(\theta_W/2)\approx 1/15$.
For small fields, say $b\lsim 0.3$, and strong transitions, 
say $\sigma/g_3^4\gsim 0.3$, this requirement is satisfied 
if the system has a linear extension $V/A > 100 g_3^{-2}$.
This distance scale is much larger than the typical correlation 
lengths of the system and thus quite difficult to achieve in practical
lattice simulations. However, it should be remembered that
even in the mixed phase, the transition takes locally 
always place between homogeneous phases, and these
are easier to simulate. 

The consideration above was at the tree-level.
To estimate the magnitude of quantum corrections, one
may compute how $H_Y$ affects the $W^\pm,Z^0$ loop contributions
in the broken phase free energy. 
At 1-loop level, the answer is
\ba
V_{W^\pm,Z^0}^{\rm 1-loop}(v,y_c)&=&-{1\over12\pi}(4m_W^3+2m_Z^3)-
{e_3Q\over2\pi}(\sqrt{m_W^2+e_3Q}-\sqrt{m_W^2-e_3Q})\nn
&&+{1\over4\pi^{3/2}}\int_0^\infty{dt\over t^{3/2}}e^{-m_W^2t}
\left({1\over t}-{e_3Q\over \sinh e_3Q t}\right),
\la{1lpot}
\ea
where $m_W=\fr12 gv, m_Z=m_W\sqrt{1+z}$, and $e_3Q=g_3'B_Y$.
Eq.~\nr{1lpot} can be reliably converted to a 1-loop change in $y_c$
if the correction is smaller than the leading magnetic energy term
following from \eq\nr{Htreel}: this requires that $h \gsim 0.15$ and
that $x$ not be too small, $1/(2\pi\sqrt{h})\ll (8x)^{3/4}$. Then,
\ba
& & y_c^\rmi{1-loop}(x,h) =  
  -h \sqrt{\frac{2x}{1+z}} + \frac{\sqrt{h}}{4\pi (8x(1+z))^{\fr14}}
  \biggl[
  \fr23+{(1+z)^{\fr32}\over3} \\
& &  + (\frac{8 x}{1+z})^{\fr12} \Bigl(
  \sqrt{1+(\frac{8x}{1+z})^{\fr12}}
  -\sqrt{1-(\frac{8x}{1+z})^{\fr12}}\Bigr) 
  -\frac{2x}{3(1+z)}
  \Bigl(1 - \frac{7}{10}\frac{x}{1+z} + {\cal O}(x^2) \Bigr) \biggr]. \nonumber
\ea
Qualitatively, the effect of 1-loop corrections is that the regime 
$x > (1+z)/8$ becomes unstable (has an imaginary part), and that the value 
of $y_c$ is somewhat less negative than at tree-level.

{\em The stability of the homogeneous phases}. According to the
standard Landau-level analysis, the effective mass squared of the Higgs
field in the symmetric phase is modified by the term $+ (g_3'/2)
B_Y^\rmi{3d}$ in the presence of a magnetic field, while the vector
mass squared remains zero. In the broken phase, in contrast, the
neutral Higgs mass squared remains unmodified, while the $W^\pm$-bosons
get a term $-e_3 Q = -g_3' B_Y^\rmi{3d}$. This term is characteristic
of non-Abelian gauge theories. For a stable phase, the mass squared
must be positive. Thus we get that the symmetric phase is stable at
\be
y > y_s = -h/2,
\ee
while the broken phase is stable at
\be
y < y_b = -4hx/(1+z).
\ee
For $y_s > y_b$, which can happen when $x > (1+z)/8$, neither phase is
stable. Therefore, in this region of the parameter space the ground state
of the system must be inhomogeneous. In a superconductor, an external
magnetic field cannot penetrate the broken Higgs phase, and the flux goes
either through a macroscopic tube of the symmetric phase (type I
superconductors) or through a vortex lattice (type II superconductors).
In the present case, the magnetic field can penetrate the broken phase.
Nevertheless, a behaviour similar to a type I or type II superconductor
can emerge. As we have seen, a type I mixed phase behaviour appears on 
the first order line shown in \fig\ref{treelevel}, for $x<(1+z)/8$.
The relevant ground state in the regime $x>(1+z)/8, y_s >
y_b$ has been studied by Ambj{\o}rn and Olesen (AO)~\cite{ao}. It
contains a $W^\pm$-condensate, with a periodic vortex-like structure,
similar to a type II superconductor. 
The emerging tree-level phase structure is illustrated in
\fig\ref{treelevel}. 

Our main aim below is to analyse numerically if this phase structure is
stable against quantum fluctuations.

\section{Magnetic fields on the lattice}
\la{lattice}

The theory in \eq\nr{contaction} can be discretized in the standard way
(see, e.g., \cite{su2u1}). The lattice action is
\ba
S&=& \beta_G \sum_x \sum_{i<j}[1-\fr12 \tr P_{ij}]
 +\fr12 \frac{\beta_G}{z} \sum_{x}\sum_{i<j} \alpha_{ij}^2  \nonumber
\\
 &-& \beta_H \sum_x \sum_{i}
\fr12\tr\Phi^\dagger(x)U_i(x)\Phi(x+i)
e^{-i\alpha_i(x)\sigma_3} \nonumber \\
 &+&
 \beta_2\sum_x\fr12\tr\Phi^\dagger(x)\Phi(x) +
 \beta_4\sum_x \bigl[ \fr12\tr\Phi^\dagger(x)\Phi(x)\bigr]^2,
\la{lattaction}
\ea
where $\alpha_i = a (g_3'/2) B_i$, $\alpha_{ij}(x) =
\alpha_i(x)+\alpha_j(x+i)
-\alpha_i(x+j)-\alpha_j(x)$ is
the discretized field strength tensor,
and $\Phi = (i\tau^2\phi^*,\phi)$. The dimensionless
lattice field is $\pdp_\rmi{latt}=\pdp_\rmi{cont}/g_3^2$,
and the couplings are
\be
\beta_G=\frac{4}{a g_3^2},\quad \beta_H=\frac{8}{\beta_G}, \quad
\beta_2=\frac{24}{\beta_G}+ \frac{64}{\beta_G^3}\frac{m^2}{g_3^4},
\quad
\beta_4 =\frac{64}{\beta_G^3} x,
\ee
where $m^2$ can be read from \eq(33) in~\cite{lc},
with $\gamma=\infty$.
In this paper, we use only one value of $\beta_G$, 
$\beta_G=8$, since we monitor mainly {\em shifts} 
in quantities whose discretization errors are already
known from simulations without a magnetic field.

On a lattice with the extent $N_1N_2N_3$,
the flux of \eq\nr{contflux} to the $x_3$-direction is imposed by
modifying the periodic boundary conditions of
the variables $\alpha_i$ as
follows
(for each fixed $n_3$ in $\alpha_i(n_1,n_2,n_3)$, not written
explicitly):
\be
g_3'\Phi_B = 2 \Bigl\{
\sum_{n_1=1}^{N_1} \Bigl[
\alpha_1(n_1,0)-\alpha_1(n_1,N_2)\Bigr] +
\sum_{n_2=1}^{N_2} \Bigl[
\alpha_2(N_1,n_2) - \alpha_2(0,n_2)
\Bigr]\Bigr\}.
\la{latticeflux}
\ee
With strictly periodic boundaries for $\alpha_i$, 
the net flux is thus zero.

In principle, any flux $\Phi_B$ can be simulated. However, the action
is not periodic unless the quantities in the square brackets in
\eq\nr{latticeflux} are integer multiples of $2\pi$. The violation
of these conditions will result in boundary defects (boundary
currents), and the lattice translational invariance will be lost. This
requirement quantizes the total flux: $g'_3\Phi_B/2 = 2\pi n$, with
$n$ an integer.  

The most economical way of implementing the boundary conditions
is to make only one $(x,y)$-plane link aperiodic.  Thus, the
boundary condition in \eq\nr{latticeflux} becomes
\be
 \alpha_1(n_1,0)-\alpha_1(n_1,N_2) =
  2n \pi \delta_{n_1,1}\,,
\ee
otherwise $\alpha_i$ periodic. Despite its appearance, this condition
does not give any special physical status to the $n_1 = 1$ -plane,
or the $x$-direction, since the `twist' can be transformed to an arbitrary
location without modifying the action.

It is essential that in \eq\nr{lattaction} the hypercharge field
$\alpha_i$ has a non-compact action. For a compact action the variables
$\alpha_i$ are only defined modulo $2\pi$, and the boundary conditions
above reduce to purely periodic ones. In this case the flux can
spontaneously fluctuate in units of $2\pi$ from configuration to
configuration, and the mean flux will average to zero.

In terms of the flux density
parameter $B_Y^\rmi{3d}$ appearing in \eq\nr{b}, the quantization
condition is $a^2 N_1 N_2 (g_3'/2) B_Y^\rmi{3d} = 2 \pi n$, and the
dimensionless variable $b$ characterizing the average magnetic 
flux density is then
\be
 b = \frac{g_3' B_Y^\rmi{3d}}{g_3^4} =
 \Bigl(\frac{\beta_G}{4}\Bigr)^2 \frac{4 \pi n}{N_1 N_2}.
\la{bquant}
\ee

As to the observables measured, let us mention that in 
addition to the usual volume averages of gauge-invariant
operators, it is useful to study 
gauge-invariant operators summed over the
$x_3$-direction, and only a small sub-block of the $(x_1,x_2)$-plane:
\be
(\pdp)_\rmi{blocked} = \sum_{n_3=1}^{N_3}
\sum_{(n_1,n_2)\in \rmi{block}} \pdp(n_1,n_2,n_3).
\la{block}
\ee
The distribution of such observables is sensitive to 
possible inhomogeneous spatial structures.

\section{Results}
\la{results}

We have performed Monte Carlo simulations with $x$ in the range
$0.1 \le x \le 0.2$ and a magnetic field in the range $0 \le b \le 24\pi/64$,
with volumes up to $64^3$ (Table \ref{statistics}).  The total
number of runs (combinations of volumes and couplings $(x,y,b)$)
is 548. As mentioned in connection with \eq\nr{finvol}, it should
be kept in mind, though, that the true infinite volume limit is 
impossible to achieve in some regions of the parameter space.

\begin{table}[t]
\centering
\begin{tabular}{lll}
\hline
$x$   & $b/(\pi/64)$ & volumes         \\ \hline
0.10  & 0, 4, 8, 12  & $16^3$, $22^3$, $28^3$, $32^3$ \\
0.11  & 24           & $32^3$ \\
0.12  & 0, 8, 16, 24 & $16^3$, $32^3$  \\
0.121 & 16           & $48^3$\\
0.125 & 24           & $16^3$, $32^3$, $48^3$  \\
0.13  & 0, 8, 16, 24 & $16^3$, $32^3$  \\
0.14  & 24           & $32^3$          \\
0.15  & 0, 8, 16, 24 & $16^3$      \\
0.16  & 24           & $32^3$, $64^3$ \\
0.20  & 0, 8, 16     & $16^3$, $32^3$  \\ \hline
\end{tabular}
\caption[a]{\protect The simulation points at fixed $x$. In each case,
several values of $y$ (up to 20) were used.  Extensive additional
simulations were performed close to the endpoints of the first order
transition lines for $b/(\pi/64) = 1$,2,3,4,6,12,16,24 (see
\fig\ref{xcb}).  All the simulations reported here are with
$\beta_G=8$.  For the volumes $22^3$ and $28^3$ the closest
approximation to the $b$ indicated,
satisfying the quantization rule in \eq\nr{bquant},
was used.}
\la{statistics}
\end{table}

\begin{figure}[t]
\centerline{\epsfxsize=8cm\epsfbox{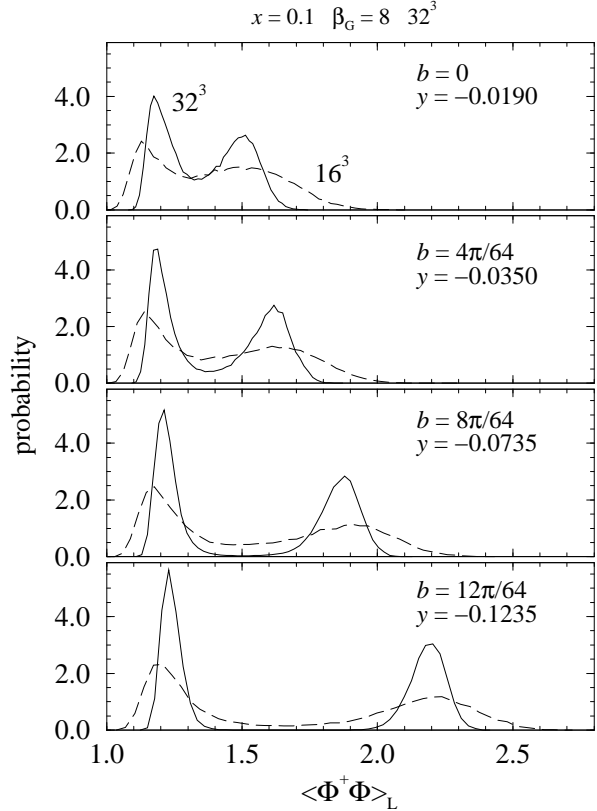}}
\caption[a]{The probability distribution of $\pdp$ at $x=0.1$ and $y=y_c$,
measured with four different flux densities and the volumes $16^3$ and
$32^3$.  The transition becomes stronger (the separation of the 
peaks increases and the probability between the peaks decreases)
and $y_c$ decreases with increasing $b$.  Simulations
with $b\ge 4\pi/64$ are multicanonical. For the cubic 
geometry employed, the ``mixed'' 
configurations where the volume average of $\pdp$ is between the
two peaks, prefer to have phase interfaces parallel with
the magnetic field. For much larger volumes, a single peak corresponding
to the mixed phase would appear in between
the two peaks seen here, see \fig\ref{x0113hg}.}  
\la{fig:x10}
\end{figure}

\begin{figure}[t]
\centerline{\epsfxsize=9cm\epsfbox{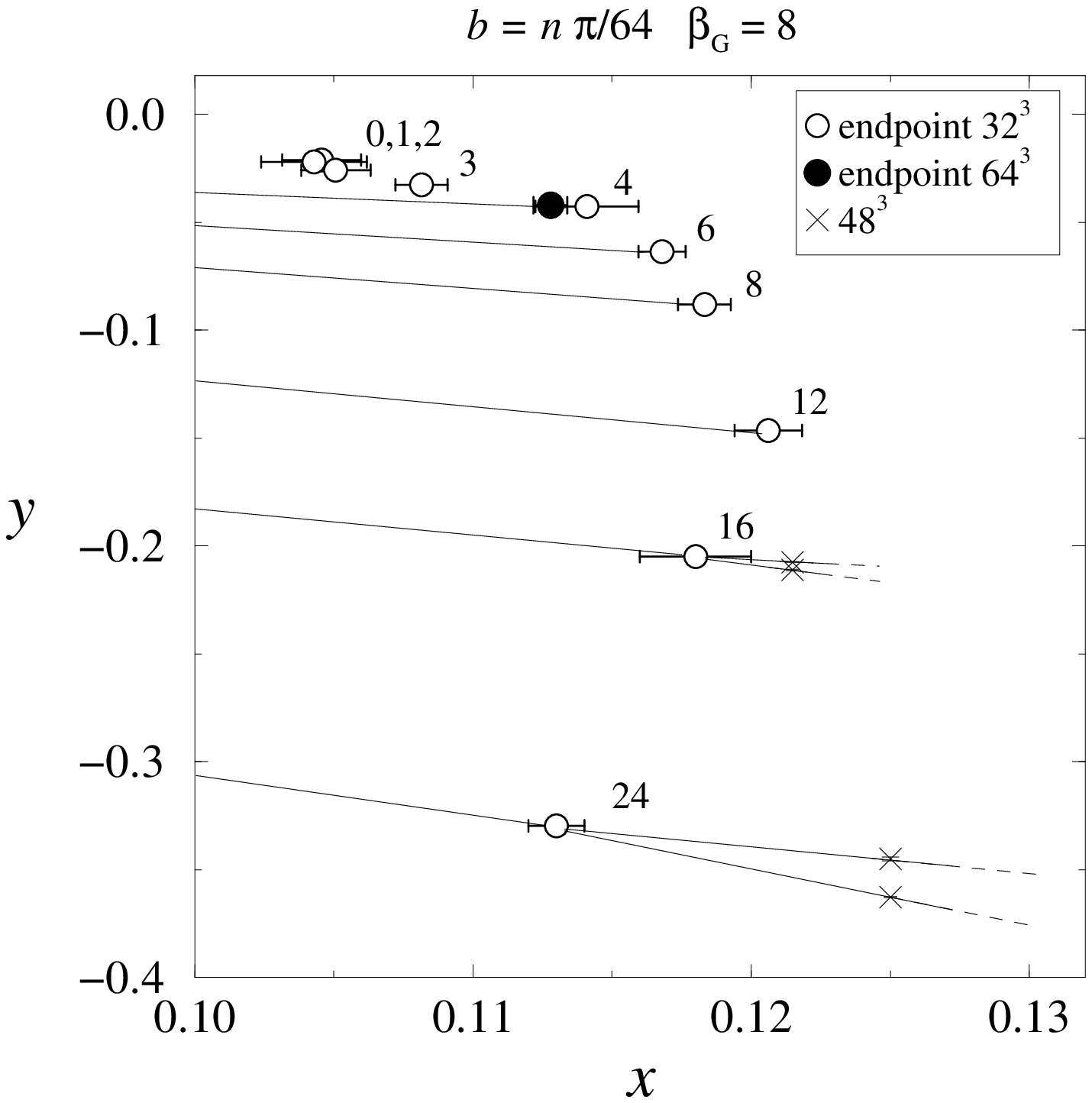}}
\caption[a]{The $0.10<x<0.13$ part of the
phase diagrams for constant $b$, $0\le b \le
24\pi/64$, and finite volumes.  
The continuous lines show first order phase transitions
(for clarity, not shown for $n<4$), and the open and filled circles
the endpoint/triple point locations.  At large fields ($n\ge 16$) a
`type I' mixed phase appears between the symmetric and broken phases.
The transitions have been located on the points shown with the crosses.  
All the transitions become weaker when $x$ increases. 
The behaviour at large $x$ is not known (dashed lines), but 
we expect the region opening at the triple point to be closed.
When the lattice size is increased, the triple points appear at
smaller $x$ and $b$.}  \la{xcb}
\end{figure}

\paragraph{Small fields.%
}
For $b=0$, the line of first order phase transitions at small $x$ has
an endpoint, in the universality class of the Ising
model~\cite{endpoint}.  
According to the discussion in Sec.~\ref{pert}, for $b \neq 0$
the transition line should split into
two lines with a mixed phase in between.  However, we find
that for small $b$ the mixed phase remains absent for all
practical lattice sizes (see \eq\nr{finvol}).
Indeed, we observe that the $b=0$ qualitative behaviour
remains there for small $b$, even though the endpoint moves to larger
$x$. Moreover, for any given $x < x_c$ (the endpoint location at
$b=0$), the transition becomes more strongly of the first order for
increasing $b$.  This is shown in \fig\ref{fig:x10} for $x=0.1$.

We have determined the finite volume
endpoint location with a method similar to that
used in~\cite{endpoint}: 
for each $b$, we locate the point $(x_c,y_c)$ where the order
parameter probability histogram $P(\pdp)$ has two peaks of equal
height, and the peaks are approximately 2.2 times higher than the
histogram height between the peaks, a characteristic value for the
magnetization distribution of the 3d Ising model at the critical
point%
\footnote{It may be noted that in ref.~\cite{endpoint}
great care was taken in order to find a good approximation for the
`magnetization' and `energy'-like observables $M$ and $E$, which map
onto the Ising-model $M$ and $E$ at the critical point.  The $M$ and
$E$-directions were found from a 6-dimensional space of lattice
operators (one of which was $\pdp$).  The endpoint was located in
terms of $P(M)$, not $P(\pdp)$.  However, it was also seen that the
probability distribution in the 6-dimensional space spanned by the
original operators is so strongly stretched along the $M$-direction
that it almost looks like a straight line (Fig.~2 in
ref.~\cite{endpoint}).  Thus, the projections of this distribution
along the $M$ and $\pdp$ -directions give distributions which have
practically the same form, and both can be used to locate the critical
point.  The $M$ and $E$-like operators were necessary for the analysis
of critical exponents in ref.~\cite{endpoint}.}.
As an example, the $b=0$
histogram shown in \fig\ref{fig:x10} almost fulfills this criterion for
$(x_c,y_c)$. (The main deviation comes from the fact that in
\fig\ref{fig:x10} the values of $y_c$ have been determined by the more
common {\em equal weight\,} criterion.  However, the equal height
method is better suited for searching for the endpoint location.)

The actual process of locating the critical point consists of two stages:
First, a few short simulations are performed near the estimated
critical point, giving a series of improved estimates.  Then, a long
simulation is performed using the best estimate for the critical point.
Finally, the resulting probability histogram $P(\pdp)$ is 
{\em reweighted\,} with respect to $x$ and $y$ until the critical point
condition mentioned above is fulfilled, which yields the final
value of $(x_c,y_c)$.  
For details of this procedure we refer to \cite{endpoint}.

The results, together with the
critical lines, are shown in \fig\ref{xcb}.  The main observation is
that $x_c$ increases quite slowly with $b$, and does not reach values
larger than $\sim 0.125$ for $b\lsim 1$.  We have used mostly $\beta_G
= 8$, volume $=32^3$ lattices for this analysis.  Strictly speaking,
the situation is slightly more complicated in the infinite volume
limit because of the appearance of the mixed phase.  Nevertheless, the
finite volume results are indicative of the {\em shift\,} of the
transition line and $(x_c,y_c)$ as functions of $b$.  Indeed, the
location of the endpoint does not change significantly when the volume
is increased from $32^3$ to $64^3$ at $b=4\pi/64$; see \fig\ref{xcb}.

\begin{figure}[t]


\centerline{ 
\epsfxsize=7.2cm\epsfbox{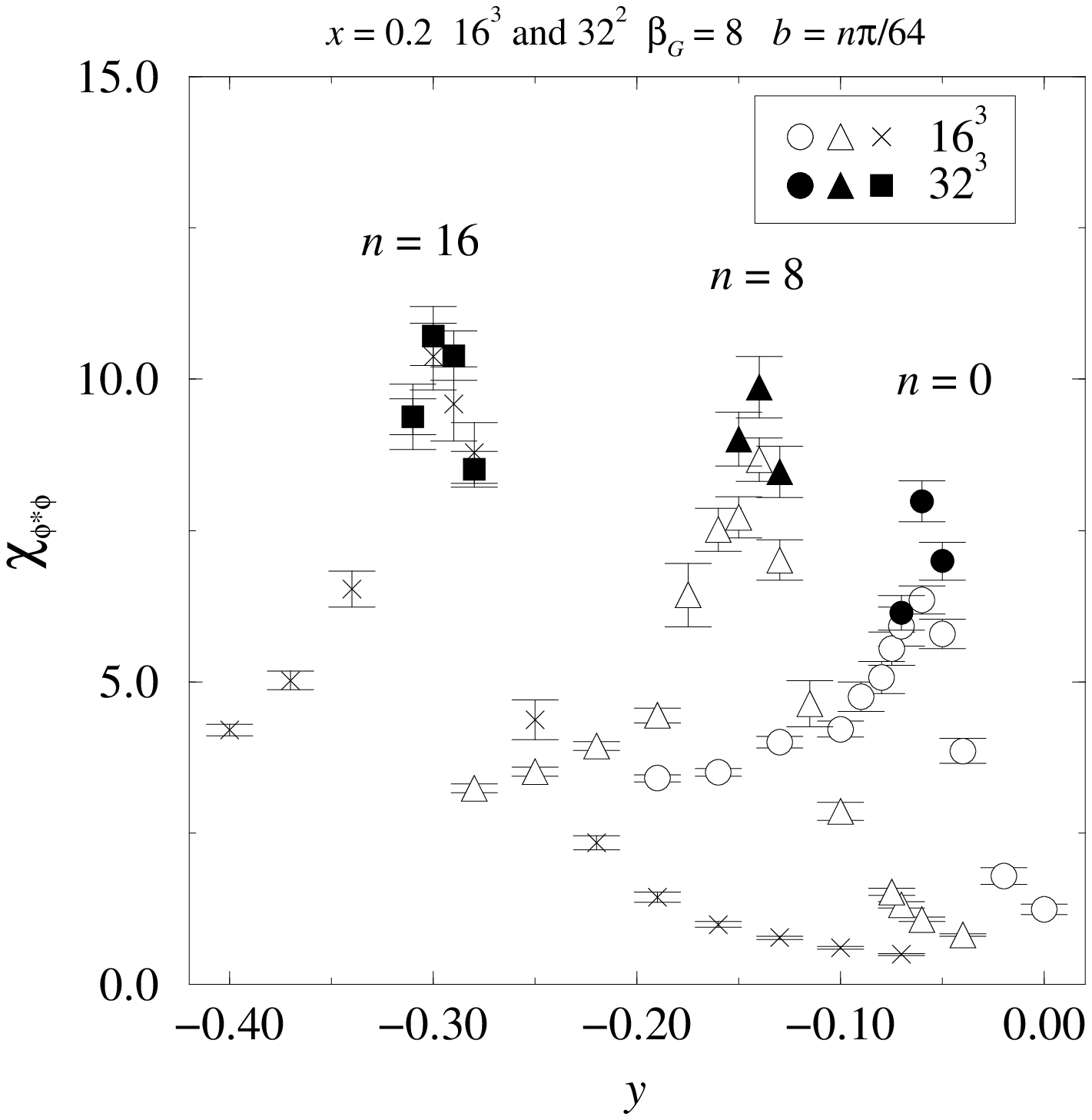} ~~~
\epsfxsize=7.5cm\epsfbox{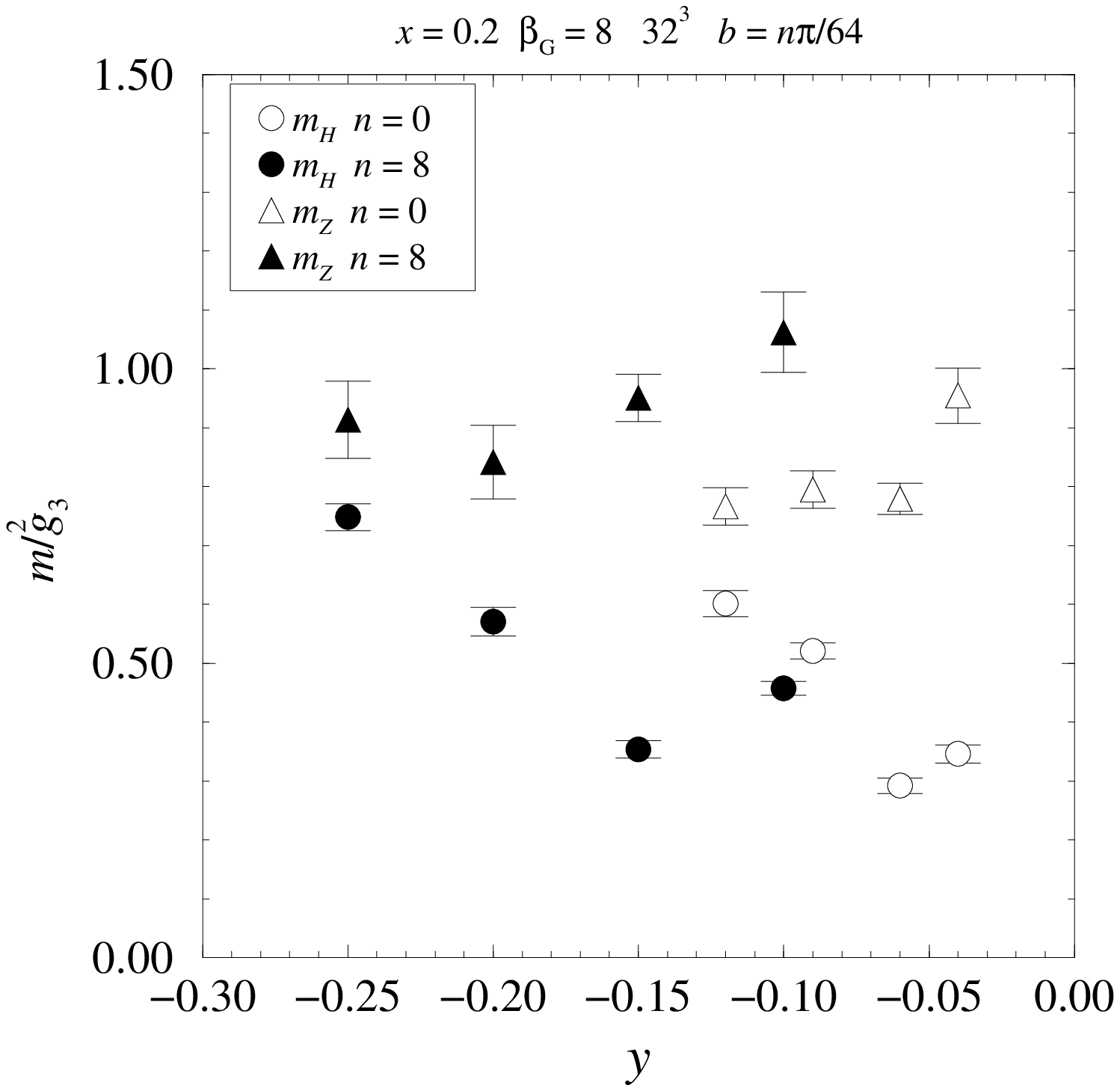}}

\caption[a]{The crossover region at $x=0.2$ for $b \le 16\pi/64$.
{\em Left:} The $\pdp$ susceptibility.
While the susceptibility has a clear peak at the crossover, it does not
diverge when the volume is increased.
{\em Right:}
The inverse correlation lengths.
In contrast with perturbation theory, there is
always a non-vanishing mass, and thus no 2nd order 
phase transition. The photon mass vanishes everywhere.}
\la{masses}
\end{figure}


\paragraph{Searching for the Ambj{\o}rn--Olesen phase.}
According to the tree-level structure in \fig\ref{treelevel}, there
should be an inhomogeneous Ambj{\o}rn--Olesen (AO) phase 
at large Higgs masses, $x>(1+z)/8$. 
In this phase $\pdp$ has a non-zero expectation value with a periodic
spatial structure in analogy with vortices in type II superconductors,
and there is a $W^\pm$-condensate.

Indeed, we {\em do\,} observe the AO phase in classical configurations
(which are solutions of the lattice equations of motion):
if we take a lattice configuration at $x>(1+z)/8$, 
$y \sim -b\sqrt{2x}$ (see \eq\nr{binter}), and `cool' it so that it minimizes
the lattice action in \eq\nr{lattaction}, a periodic $\phi$ and $W$ 
structure emerges.
However, the `quantum' configurations (which contribute to the lattice
partition function) contain a lot of fluctuations, and it is far
from straightforward to observe whether 
there is an underlying structure in them.

The cooling mentioned
proceeds very much like the standard lattice Monte Carlo
update, but instead of a new thermally distributed value, the field
variables are chosen so that the local action is minimized.  In one
cooling sweep the action is minimized at each lattice point and for
all of the lattice fields.  The UV noise vanishes after only a few
cooling sweeps, but cooling down to the classical configuration
can take several thousands of update sweeps.

Using the standard global thermodynamic quantities on
the lattice, on the other hand, 
we {\em do not} observe any transition to a phase with
qualitatively new properties at $x=0.2$.  The behaviour of the $\pdp$
susceptibility $\chi_{\pdp} = V \langle (\pdp - \langle\pdp\rangle)^2 \rangle$
is shown in \fig\ref{masses}(left), and the behaviour
does not change qualitatively from that at $b=0$.  

It is also possible to directly try to measure the $W^+W^-$-condensate
on the lattice.  For example, the (lattice) operator $W^+_1 W^-_2 -
W^+_2 W^-_1$ is gauge invariant and, in the absence of an external
magnetic field, has a vanishing expectation value.  However, when
there is an external field in the $x_3$-direction, this operator
acquires a non-zero expectation value, even deep in the symmetric and
broken phases where there is no $W^+ W^-$-condensate.  While we do
observe a clear increase of the condensate operator in the crossover
region, a similar increase is also seen near the first order
transition.  Thus, we cannot assign this increase to the appearance of
a $W^+ W^-$-condensate.

A clear signature of the AO phase is its periodic spatial structure
in the $(x_1,x_2)$-plane (or, if the strict periodicity is
destroyed by fluctuations, there should at least be an enhancement of the
characteristic wavelengths of the lattice fields).  We have attempted
to directly observe this on the lattice at $x=0.2$, $b \le 16\pi/64$
with various methods:

\begin{itemize}
\item
A moderate amount of
cooling is a very efficient method of getting rid of the lattice UV
noise.  However, in this case the interpretation of the
results is far from clear:
if we cool the configuration too much, it will be driven towards the
solution of the classical equations of motion and finally reveal the
structure of the AO phase.  To deem the `correct' amount of
cooling is ambiguous, at best.  It is much more objective to
measure the properties of configurations without any cooling.
\item
The Fourier power spectrum of $\pdp$ in the $(x_1,x_2)$-plane should
reveal the presence of periodic spatial structures.  However, in our
measurements the spectrum remains qualitatively 
the same for both $b=0$
and $b\ne 0$.
\item
Since the AO field structure is of classical nature, it should persist
on the lattice for several update cycles, whereas the quantum noise
rapidly averages to zero.  This can be utilized by averaging the
fields over a various number of consecutive configurations (and over
the $x_3$-direction).  However, this analysis does not reveal any
non-trivial structure.
\item
All polarizations of the photon, in all directions, remain
massless.
\end{itemize}
We thus have to conclude that small magnetic fields do not result in an
inhomogeneous phase at around $x=0.2$.
A possible explanation is that, in contrast to tree-level
perturbation theory, the non-perturbative inverse correlation lengths
are always non-vanishing in this regime, as shown in \fig\ref{masses}(right).
Thus there may be some minimum value of $b$ which is needed for the 
instability and the inhomogeneous phase to appear.

\begin{figure}[tb]

\vspace*{-1cm}

\centerline{\epsfxsize=13cm\epsfbox{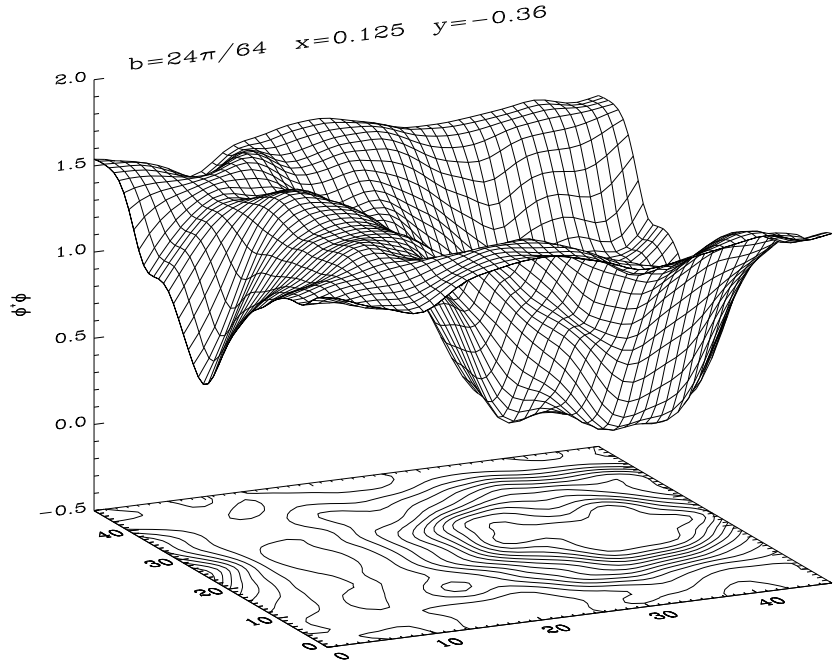}}
\caption[a]{A Higgs field configuration on a $48^3$ lattice
for $x=0.125$, $y=-0.36$, just above the transition to the broken
homogeneous phase.  A thermodynamically {\em stable\,} cylinder of 
the symmetric
phase is surrounded by the broken phase.  In order to reduce noise in
the plot, we have cooled the configuration with 10 local cooling
sweeps, and averaged the fields over the $x_3$-direction.}  \la{typeI}
\end{figure}

\begin{figure}[tb]
\centerline{\epsfxsize=9cm\epsfbox{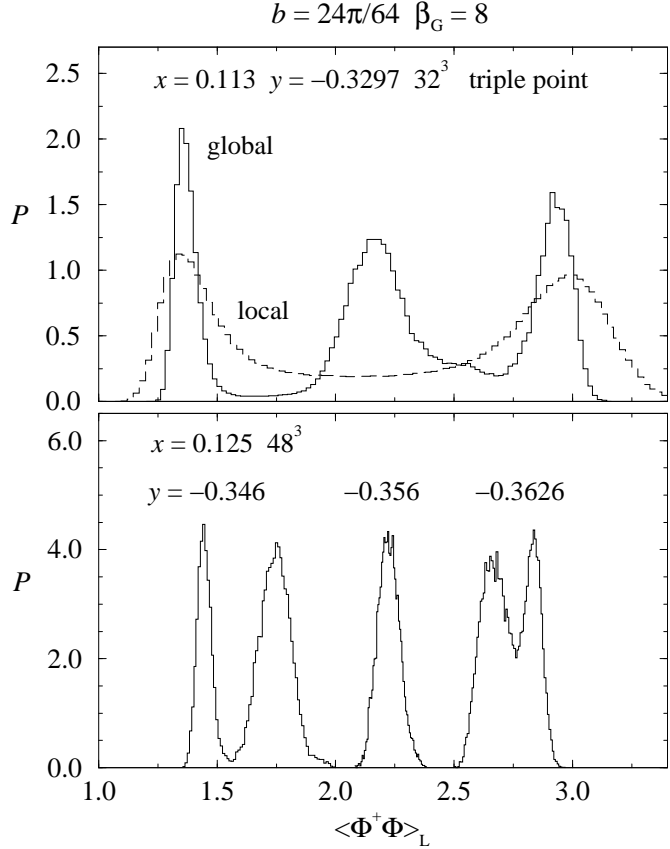}}
\caption[a]{ The appearance of the `type I' mixed phase at $b=24\pi/64$.
{\em Top:} The $\pdp$ histogram measured at the triple point, where the
symmetric phase (left peak), broken phase (right peak) and the
mixed phase (center) coexist.  Strong first order transitions separate
the phases at this finite volume.  
A dashed line shows the distribution of the blocked $\pdp$
in the mixed phase,
measured over $1/16$th 
of the total area (see \eq\nr{block}).  
Here the center peak is absent, in accord with the
mixed broken/symmetric spatial structure of the mixed phase.  
{\em Bottom:} The global $\pdp$ histograms at $x=0.125$, $V=48^3$.  
Separate first order
transitions occur at $y\approx-0.346$ and $y\approx-0.3626$; for
intermediate values of $y$ the mixed phase is stable.}  \la{x0113hg}
\end{figure}

\paragraph{Large fields.
}
For large values of the magnetic field we do observe a mixed phase. 
This phase 
consists of a domain of the symmetric phase surrounded by the broken phase.
A moderately
cooled Higgs field configuration\footnote{In this case the existence of the
mixed phase is clear even without cooling (see below); 
the cooling makes the figure only
much easier to read.} is shown in \fig\ref{typeI}, for
$x=0.125$, $y=-0.36$, and $b=24\pi/64$.
Inside the symmetric phase domain, the hypermagnetic field is
somewhat ($\sim 25\%$) larger than elsewhere.  For the
Higgs field, the behaviour is similar to that in a type I
superconductor, and we shall thus
use the name ``type I'' for this mixed phase. 

For each lattice size and (large enough) $b$, the type I phase appears
only in some specific interval of $x$: if $x$ is too small, the first
order nature of the symmetric $\leftrightarrow$ broken transition is
too strong for the mixed phase to appear in the volumes we can
simulate in practice.  This is due to the tension (free energy density) 
of the interface between the symmetric and broken domains
in the mixed phase.  At a {\em triple point\,} $(x,y)_{\rm triple}$,
the symmetric, broken and mixed phases are equally likely to
appear. This is illustrated in the top part of \fig\ref{x0113hg}.  The
transitions separating the three phases are of the first order.  When
$x$ is further increased, separate symmetric $\leftrightarrow$ type I
and type I $\leftrightarrow$ broken transitions appear.  This is shown
in the bottom part of \fig\ref{x0113hg}.

The resulting phase diagrams are shown in \fig\ref{xcb}.  When $x >
x_{\rm triple}$, a band of the type~I phase appears, separated by lines of
first order transitions.  For each fixed $(x,y)$, the type I mixed phase
contains a definite volume fraction of the symmetric phase
(by a volume fraction 
we mean the weight of the corresponding peak in the
local 'blocked' distribution, see \fig\ref{x0113hg}).
At the upper (lower) critical $y_{c}(x)$, the symmetric phase occupies
the largest (smallest) volume fraction.  At the triple point these
limits are equal.  For a finite total volume, 
it is not possible to have a mixed phase
with arbitrary volume fractions.

When the volume is increased, the interface tension becomes less
significant in comparison with the bulk free energies.  Thus, when
the volume goes to infinity, $x_{\rm triple}\rightarrow 0$, 
and the mixed phase band becomes slightly wider (since now 
it is possible to have arbitrary volume fractions).  However, the
simulations near the triple point are very difficult, and we have not
attempted to study the volume dependence of $x_{\rm triple}$
in detail.

The first order transitions 
separating the homogeneous phases from the mixed phase 
become weaker when $x$ increases.  At
$x \approx 0.2$ we have not observed transitions any more, and
symmetric and broken phases appear to be analytically connected.
However, since the type I phase breaks translational invariance
(in the sense that after the removal of a zero mode, 
the observables measured depend on the location), it
has to be separated from the symmetric and broken phases by some kind
of phase transitions.  Thus, the type I regions in \fig\ref{xcb} should
form isolated domains, or they could also transform into an AO
phase at some $x$.

Finally, 
a couple of technical issues about the phase structure on the lattice:

First, the locations of the transitions between the type I and the
homogeneous phases are very sensitive to finite size effects.  The volume
should be large enough in order for the phase interfaces in the mixed
phase to have a negligible effect.  This is
very difficult to achieve in practice.  In \fig\ref{xcb} the transitions
have been located with a $48^3$ lattice; using a larger volume would
widen the type I region, and move the triple point to smaller $x$.

Second, on a periodic toroidal lattice there are also two other
transitions besides the ones shown in Figs.~\ref{xcb}, \ref{x0113hg}:
these correspond to transitions where a cylindrical domain
(\fig\ref{typeI}) becomes a slab spanning the lattice in $(x_1,x_3)$ or
$(x_2,x_3)$-directions.  These transitions are not physical, since they
are caused by the boundary conditions.

\section{Conclusions} 
\la{conclusions}

We have found that for Higgs mass $m_H\gsim 80$ GeV,
even magnetic fields up to $H_Y/T^2 \sim 0.3$ ($b\sim 0.6$) 
do not suffice to make
the transition be of the first order: there is only a crossover. This
is in contrast to the perturbative estimates in~\cite{gs,eek}.
Moreover, we do not observe any sign of the exotic phase with broken
translational invariance proposed by Ambj{\o}rn and Olesen for these
magnetic fields: all the gauge-invariant operators and correlation
lengths we have studied behave qualitatively as without a magnetic
field, even though the solution of the classical equations of motion
has a vortex structure with a $W^\pm$-condensate. We conclude that
fluctuations are strong enough to remove the non-trivial structure
for the parameter values studied.

On the other hand, for the finite volumes studied, we 
do observe the emergence of a new phase when the magnetic field 
is increased above $H_Y/T^2 \sim 0.3$. This phase is not of
the type proposed by Ambj{\o}rn and Olesen, though, but it is
a mixed phase, with a Higgs
field distribution similar to a type I superconductor. It would be
interesting to determine whether this
phase turns into the Ambj{\o}rn-Olesen
phase at larger Higgs masses. We expect that the phases with 
broken translational invariance (i.e., type I and AO), appear in a 
closed region of the parameter space, terminating at some $x_c$.
The fact that the transition terminates also at a non-vanishing
magnetic field, is a non-perturbative phenomenon
and does not appear in the
tree-level phase diagram in \fig\ref{treelevel}.
A more precise study of these questions is in progress.

The implications of a magnetically stabilized
mixed phase structure on different scenarios of
electroweak baryogenesis have, to our knowledge, 
not been studied before, but there might be some effects.  
A precise investigation
of these issues is beyond the scope of the present paper. 
 
\section*{Acknowledgements}

The simulations were carried out with a Cray T3E at the Center for
Scientific Computing, Finland.  We thank J. Ambj{\o}rn, K. Kainulainen,
P. Olesen, A. Rajantie and M. Tsypin for very useful discussions. This 
work was partly supported by the TMR network {\em Finite Temperature Phase
Transitions in Particle Physics}, EU contract no.\ FMRX-CT97-0122.

\end{document}